\documentclass[reprint, prd, eqsecnum, amsmath, amsfonts, nofootinbib]{revtex4-2}

\usepackage{physics}
\usepackage[colorlinks=true, citecolor=blue]{hyperref}

\newcommand{\pdd}{\partial}

\newcommand{\mc}{\mathcal}

\newcommand{\bbR}{\mathbb{R}}

\newcommand{\ovl}{\overline}
\newcommand{\eps}{\epsilon}
\newcommand{\ttb}{T\overline T}
\newcommand{\zb}{{\bar z}}

\begin{document}
\title{$\ttb$-Deformations of Holographic Warped CFTs}
\author{Rahul Poddar}
\affiliation{Science Institute, University of Iceland,\\Dunhaga 3, 107 Reykjav\'ik, Iceland}
\email{rap19@hi.is}

\begin{abstract}
  We explore $\ttb$ deformations of Warped Conformal Field Theories (WCFTs) in two dimensions as examples of $\ttb$ deformed non-relativistic quantum field theories.
  WCFTs are quantum field theories with a Virasoro$\times$U(1) Kac-Moody symmetry.
  We compute the deformed symmetry algebra of a $\ttb$ deformed holographic WCFT, using the asymptotic symmetries of AdS$_3$ with $\ttb$ deformed Comp\'ere, Song and Strominger (CSS) boundary conditions.
  The U(1) Kac-Moody symmetry survives provided one allows the boundary metric to transform under the asymptotic symmetry. 
  The Virasoro sector remains but is now deformed and no longer chiral.
\end{abstract}

\maketitle

\section{Introduction}

A Warped Conformal Field Theory (WCFT) is a quantum field theory with an SL(2,$\bbR)\times$U(1) global symmetry in two dimensions, which breaks Lorentz invariance.
Such QFTs have translation invariance, but scaling invariance is restricted to only one coordinate. 
Finite warped symmetry transformations take the form \cite{Hofman:2011zj,Detournay:2012pc}
\begin{equation}
  \label{eq:WarpedTransf}
  z \to f(z),\quad \zb \to \zb + g(z). 
\end{equation}
However, despite not being Lorentz invariant, this class of two-dimensional quantum field theories still possesses an infinite-dimensional symmetry algebra, namely a Virasoro$\times$U(1) Kac-Moody current algebra.
WCFTs are interesting as they appear in a number of holographic systems with an SL(2,$\bbR)\times$U(1) symmetry, such as Warped AdS$_3$ \cite{Hofman:2014loa}, the near horizon geometry of extremal rotating black holes \cite{Bardeen:1999px,Dias:2007nj}, and AdS$_3$ with Dirichlet-Neumann boundary conditions \cite{Compere:2013bya}.
Holographic WCFTs have passed a number of consistency checks, such as a Cardy formula \cite{Detournay:2012pc}, holographic entanglement entropy \cite{Anninos:2013nja,Castro:2015csg,Song:2016gtd,Apolo:2020qjm} and one-loop determinants \cite{Castro:2017mfj}. 

Since WCFTs are non-relativistic, they do not couple to standard (pseudo-)Riemannian manifolds.
One approach is to couple WCFTs to ``warped geometries'' \cite{Hofman:2014loa}, a variant of Newton-Cartan geometries.
These geometries can be found at the boundary of Warped AdS$_3$ spacetimes.
Unfortunately, these geometries have certain pathologies, such as a degenerate metric, which make some calculations untenable.
Another way to couple WCFT to a background manifold is to allow the manifold to transform with the warped symmetry transformations.
Holographically, this requires relaxing Dirichlet boundary conditions of the bulk metric to  boundary conditions which allow for asymptotic symmetry transformations to transform the boundary metric under the warped symmetry transformation of the boundary WCFT.
The Dirichlet-Neumann boundary conditions of Comp\'ere, Song and Strominger (CSS) \cite{Compere:2013bya} do exactly this. 
This approach bypasses the need for a warped geometry with degenerate metrics, and we can work with conventional techniques.
 
Two dimensional translationally invariant quantum field theories admit a class of solvable irrelevant deformations built from conserved currents, most notable of which is the $\ttb$ deformation \cite{Zamolodchikov:2004ce, Smirnov:2016lqw}.
The $\ttb$ operator is defined by the determinant of the energy-momentum tensor of the quantum field theory, and the deformed action obeys the following flow equation
\begin{equation}    
  \label{eq:ttbardef}
  \begin{split}
    \pdd_\lambda S_{\text{QFT}}(\lambda) &= -\frac12\int d^2x \sqrt{\gamma}\,\mc O_{\ttb}^{(\lambda)},\\
    \mc O_{\ttb} &= \det T = \frac12\eps^{\mu\rho} \eps^{\nu\sigma} T_{\mu\nu} T_{\rho\sigma},
  \end{split}
\end{equation}
where the deformation parameter $\lambda$ is the coupling to the $\ttb$ operator $\mc O_{\ttb}$. 
This operator is defined using point splitting, which in the coincident limit produces a well defined local operator up to total derivatives.
The expectation value of $\mc O_{\ttb}$ turns out to be a constant, and from this one can derive the flow of energy eigenstates of the quantum field theory defined on a cylinder of radius $R$,
\begin{equation}
  \label{eq:burger}
  \pdv{E_n}{\lambda} = E_n \pdv{E_n}{R}+\frac{P_n^2}{R}. 
\end{equation}
Even though the energy eigenvalues are changed, the Hilbert space remains undeformed since there is a one-one correspondence between the states of the original and deformed theory. 
Similarly, other observables can be calculated in the deformed theory, for example the deformed Lagrangian, partition function, two-two scattering matrices, correlation functions, etc \cite{Caselle:2013dra,Cavaglia:2016oda,Dubovsky:2013ira,Dubovsky:2018bmo,Cardy:2018sdv, Cardy:2019qao, Aharony:2023dod,Benjamin:2023nts}.

There are various ways to interpret how the $\ttb$ deformation acts on holographic CFTs.
One proposal by \cite{McGough:2016lol} is to impose Dirichlet boundary conditions for the bulk metric at a finite radius.
Another proposal given by \cite{Guica:2019nzm} is to treat the $\ttb$ deformation as a double trace deformation, which will deform the asymptotic behaviour of the bulk fields \cite{Klebanov:1999tb,Witten:2001ua}.
This approach agrees with the cut-off AdS proposal when both are valid, but has the advantage of working when there are bulk matter fields, and also for either sign of the deformation parameter, which the former does not.
More recently there has also been the ``Glue-on AdS holography'' proposal \cite{Apolo:2023vnm} which also agrees with \cite{Guica:2019nzm} for the positive sign of the deformation parameter in the absence of matter fields. 

Using the mixed boundary conditions/Dirichlet boundary conditions at finite radius, the asymptotic symmetry algebra of the bulk dual to $\ttb$ deformed holographic CFT was calculated in \cite{Guica:2019nzm,Kraus:2021cwf,He:2021bhj}.
Despite losing conformal invariance, the asymptotic symmetry algebra turns out to still have a Virasoro$\times$Virasoro structure.
However either the central charge becomes state dependent, or one loses the holomorphic factorization of the symmetry algebra, which can also be expressed as a non-linear deformation of the standard Virasoro algebra. 

In this work, we explore $\ttb$ deformations of WCFTs from a holographic perspective.
To establish what a $\ttb$ deformation of a WCFT is, one must first define what the energy-momentum tensor of a WCFT is.
Canonically, for WCFTs defined on a warped geometry, energy-momentum tensors are not symmetric and a determinant is harder to define since the metric is degenerate and non-invertible.
The energy-momentum tensor turns out to be a tensor with components being a chiral stress tensor and a U(1) current. 
For WCFTs dual to Warped AdS$_3$, it is also not possible to use the Fefferman-Graham expansion to compute the energy momentum tensor for the same reason, the boundary metric is not invertible.
However, if we study WCFTs dual to AdS$_3$ with CSS boundary conditions, for the price of a boundary metric which is not invariant under warped transformations, we have an invertible metric and a symmetric energy-momentum tensor, and a conventional definition for a determinant.
Given these considerations, it is possible to propose a definition for a $\ttb$ deformed WCFT which is dual to AdS$_3$ with $\ttb$ deformed CSS boundary conditions.

This paper is organized as follows.
We first briefly review the mixed boundary conditions of \cite{Guica:2019nzm} in Section \ref{sec:ttbrev}.
Then in Section \ref{sec:css} we review the CSS boundary conditions and derive the Virasoro$\times$U(1) Kac-Moody algebra.
In Section \ref{sec:defcss} we derive the $\ttb$ deformed CSS boundary conditions to compute the deformed symmetry algebra of a $\ttb$ deformed WCFT.
We will see that if one imposes deformed boundary conditions equivalent to Dirichlet boundary conditions at the radial cut-off surface, we recover a deformed Virasoro algebra, but we lose the U(1) Kac-Moody algebra.
However, if we allow the Dirichlet-Neumann boundary conditions to remain at the cut-off surface, which is what the mixed boundary conditions suggests is the correct approach, we recover an undeformed Kac-Moody symmetry.
We then conclude and discuss future directions in Section \ref{sec:disc}. 

\section{Review}
\label{sec:rev}
\subsection{Mixed Boundary Conditions from $\ttb$}
\label{sec:ttbrev}

We begin by briefly reviewing the mixed boundary conditions derived in \cite{Guica:2019nzm} from the variational principle.
The variation of the boundary QFT action with respect to the boundary metric sources the energy-momentum tensor of the QFT and of the bulk dual.
The flow of the variation of the QFT action is equal to the variation of the deformation which generates the flow.
So we have
\begin{equation}
  \begin{split}
    \pdd_\lambda \delta S &= \delta \pdd_\lambda S\\
    \pdd_\lambda \left( \frac12 \int_{\pdd \mc M} d^2 x \sqrt{\gamma} T^{(\lambda)}_{ij}\delta\gamma^{(\lambda)ij}\right)
    &= \delta \int_{\pdd \mc M} d^2 x \sqrt{\gamma} \mc O^{(\lambda)}_{\ttb}.
  \end{split}
\end{equation}
From this we can compute the flow equations for the boundary metric and the energy-momentum tensor with respect to the deformation parameter $\lambda$.
Expressing the equations in terms of the trace reversed energy-momentum tensor $\hat T_{ij}= T_{ij} - \gamma_{ij} T^i_i $, we have:
\begin{equation}
  \label{eq:floweqn}
  \begin{split}
    \pdd_\lambda&\gamma_{ij} = -2\hat T_{ij},\\
    \pdd_\lambda&\hat T_{ij}=- \hat T_{il} \hat T_j^{~l},\\
    \pdd_\lambda&(\hat T_{il} \hat T_j^{~l})=0.
  \end{split}
\end{equation}
Solving these equations, we can express the deformed metric and energy-momentum tensor in terms of the undeformed metric and energy-momentum tensor,
\begin{equation}
  \label{eq:flowsol}
  \begin{split}
    \gamma_{ij}(\lambda)&=\gamma_{ij} - 2 \lambda \hat T_{ij} +\lambda^2 \hat T_{ik} \hat T_{jl}\gamma^{kl},\\
    \hat T_{ij}(\lambda)&= \hat T_{ij} - \lambda \hat T_{ik} \hat T_{jl}\gamma^{kl}, 
  \end{split}
\end{equation}
where everything on the right hand side are undeformed quantities.
The new deformed quantities are now the new boundary conditions for the bulk fields.
To see this, let us consider pure Einstein gravity. 

For pure Einstein gravity in three dimensions, the Fefferman-Graham expansion of the metric truncates at 2nd order in $1/r^2$ \cite{Skenderis:1999nb}
\begin{equation}
  \label{eq:FGgauge}   
  \begin{split}
    ds^2 &= l^2 \frac{dr^2}{r^2}+g_{ab}dz^adz^b\\
         &= l^2 \frac{dr^2}{r^2}+l^2r^2\left(g_{ab}^{(0)}+\frac{g_{ab}^{(2)}}{r^2}+\frac{g_{ab}^{(4)}}{r^4}\right)dz^a dz^b, 
  \end{split}
\end{equation}
using which we can now express the boundary energy-momentum tensor in terms of the Fefferman-Graham expansion
\begin{equation}
  \hat T_{ab} = \frac{k}{2\pi} g_{ab}^{(2)},
\end{equation}
where $k=\frac{l}{4G_N}$. 
For pure gravity, we also have
\begin{equation}
  g_{ab}^{(4)}=\frac14 g^{(2)}_{ac}g^{(2)}_{db}g_{(0)}^{cd}.
\end{equation}
Therefore we can express the deformed boundary metric and energy-momentum tensor in terms of the Fefferman-Graham expansion
\begin{equation}
  \label{eq:defrmfg}
  \begin{split}
    \gamma_{ab}(\lambda) &= l^2 \left(g_{ab}^{(0)}- \left(2\lambda\frac{k}{2\pi}\right)g_{ab}^{(2)}+ \left(2\lambda\frac{k}{2\pi}\right)^2 g_{ab}^{(4)}\right),\\
    \hat T_{ab}(\lambda) &= \frac{k}{2\pi}\left(g_{ab}^{(2)}-\left(2\lambda\frac{k}{2\pi}\right) g_{ab}^{(4)} \right).
  \end{split}
\end{equation}
Equating this to the Fefferman-Graham expansion \eqref{eq:FGgauge}, it is easy to see that the deformed boundary metric can be thought of as being placed at a finite radius $r_c = \sqrt{-\frac{\pi}{k\lambda}}$.
Indeed, it turns out that the Brown-York energy-momentum tensor (with the appropriate counterterm) evaluated at this surface reproduces the deformed energy-momentum tensor derived here. 
This makes it clear that in pure gravity, the mixed boundary conditions and imposing Dirichlet boundary conditions at $r_c = \sqrt{-\frac{\pi}{k\lambda}}$ are equivalent\footnote{We will be absorbing the factor of $\pi$ into the normalization of $\lambda$ and $\mc O_{\ttb}$ from now on to avoid clutter in the equations.}.

For a derivation of the mixed boundary conditions from the Chern-Simons formulation of 3d gravity, see \cite{Llabres:2019jtx}.

\subsection{CSS boundary conditions}
\label{sec:css}
Examples of constructing a holographic bulk dual to a WCFT are either Warped AdS$_3$ or AdS$_3$ with CSS boundary conditions.
We shall use the CSS boundary conditions since it is amenable to the mixed boundary conditions from the $\ttb$ deformation.

Expressing the metric in Fefferman-Graham gauge \eqref{eq:FGgauge}, we have the the following Dirichlet-Neumann boundary conditions for the metric \cite{Compere:2013bya}
\begin{equation}
  \label{eq:CSSbdryconds}
  g^{(0)}=\begin{pmatrix} P'(z)&-\frac12\\-\frac12&0 \end{pmatrix},\quad    
  g^{(2)}_{\zb\zb}=\frac{\Delta}{k},
\end{equation}
where $k = \frac{l}{4G_N}$, and $\Delta$ is a constant.
These fall off conditions are chiral, with $P(z)$ being an undetermined holomorphic function. 
This is to accommodate \eqref{eq:WarpedTransf}, which shifts $P'(z)$, and hence we must leave it undetermined. 
Note that this is unlike the warped geometry in \cite{Hofman:2014loa}, where the warped geometry metric is invariant under \eqref{eq:WarpedTransf}.

One can compute the full bulk metric with the CSS boundary conditions by taking the Fefferman-Graham expansion \eqref{eq:FGgauge} to be
\begin{equation}
  \label{eq:CSSmetric}
  \begin{split}
    &\frac{ds^2}{l^2} = \frac{dr^2}{r^2}+ \frac{\Delta}{k}d\zb^2 - \left(r^2+\frac{2\Delta P'(z)}{k} + \frac{\Delta L(z)}{k^2 r^2}\right)dzd\zb\\
    &+ \left(r^2 P'(z) + \frac{(L(z)+\Delta P'(z))^2}{k} + \frac{\Delta L(z) P'(z)}{k^2r^2}\right) dz^2.
  \end{split}
\end{equation}
Here both $L(z)$ and $P(z)$ are undetermined holomorphic functions, and parametrize the phase space of AdS$_3$ with CSS boundary conditions. 
A special case is the BTZ black hole when $P'(z) = L'(z) = 0$ \cite{Banados:1992wn,Compere:2013bya}.

The asymptotic symmetries of this metric are interesting as they differ from the usual product of $SL(2,\bbR)$ algebras despite being locally AdS$_3$.
To compute asymptotic Killing vectors, we first require that they preserves radial gauge,
\begin{equation}
  \label{eq:radialgauge}
  \mc L_\xi g_{r\mu}=0.       
\end{equation}
This fixes the asymptotic Killing vector $\xi$ to have the form
\begin{equation}
  \label{eq:radgaugexi}
  \xi = r f(z,\zb) \pdd_r +\left(V^a(z,\zb) - \int\frac{g^{ab}}{r}\pdd_bf(z,\zb) dr \right)\pdd_a. 
\end{equation}
Evaluating this for the CSS metric \eqref{eq:CSSmetric} we get, 
\begin{equation}
  \begin{split}
    &\xi^r = r f(z,\zb),\\
    &\xi^z = V^z(z,\zb) - \frac{k\big(\pdd_zf(z,\zb)+(k r^2 + \Delta P'(z))\pdd_\zb f(z,\zb)\big)}{k^2r^4-\Delta L(z)},\\
    &\xi^\zb = V^{\zb}(z,\zb) - \frac{k}{k^2r^4-\Delta L(z)}\Big((k r^2+\Delta P'(z))\pdd_zf(z,\zb)\\    
    &\qquad+((2kr^2 + \Delta P'(z))P'(z)+L(z))\pdd_\zb f(z,\zb)\Big).
  \end{split}
\end{equation}
If we impose Dirichlet boundary conditions at infinity
\begin{equation}
  \label{eq:diricinf}
  \lim_{r\to \infty}\mc L_{\xi}g_{\mu\nu}=0, 
\end{equation}
we get conditions on the undetermined functions in $\xi$:
\begin{equation}
  \begin{split}
  &\pdd_\zb V^a(z,\zb)=0,\; f(z,\zb)=-\frac12 \pdd_z V^z(z,\zb),\\
  &V^\zb(z,\zb) = P'(z) V^z(z,\zb),
  \end{split}
\end{equation}
and so we can write our asymptotic killing vector, (where $V(z)\equiv V^z(z,\zb)$)
\begin{equation}  
  \label{eq:CSSxiL}
  \begin{split}
    \xi(V) &= -\frac12 V'(z)\pdd_r +\left(V(z)+\frac{k\Delta V''(z)}{2(k^2r^4-\Delta L(z))}\right)\pdd_z\\
    &\quad+ \left(P'(z)V(z) +\frac{k(kr^2+\Delta P'(z))}{2(k^2r^4-\Delta L(z))}V''(z)\right)\pdd_\zb.
  \end{split}
\end{equation}

Asymptotic Killing vectors generate flow in the phase space, i.e.
\begin{equation}
  \mc L_\xi g_{\mu,\nu}= \pdd_{L(z)}g_{\mu\nu} \delta_\xi L(z)+ \pdd_{P'(z)}g_{\mu\nu}\delta_\xi P'(z). 
\end{equation}
From this, we can compute $\delta L$ and $\delta P$. 
It turns out that $\xi$ only transforms $L(z)$, and reproduces the infinitesimal Schwarzian transformation.
\begin{equation}
  \label{eq:dCSSxi}
  \delta_\xi L(z) = V(z)L'(z)+2V'(z)L(z)-\frac{k}{2}V'''(z),\; \delta_\xi P = 0.
\end{equation}

To transform $P(z)$, we can not allow the asymptotic killing vector to satisfy Dirichlet boundary conditions at infinity \eqref{eq:diricinf}, since warped symmetry requires changing the boundary metric.
The ``asymptotic Killing vector''\footnote{The quotes are to indicate that since this vector does not satisfy Dirichlet boundary conditions, it is technically not an asymptotic Killing vector, but since it generates flows in the phase space it will continue to be referred to as such later in this paper.} which generates transformations in $P(z)$ is
\begin{equation}
  \label{eq:CSSxiP}
  \eta(\sigma) = \sigma(z)\pdd_\zb,
\end{equation}
and the transformations of the parametrizing functions are
\begin{equation}
  \label{eq:dCSSeta}
  \delta_\eta L = 0,\quad \delta_\eta P(z) = -\sigma(z). 
\end{equation}
Note that $\eta$ also generates the warped symmetry transformation $\zb\to \zb+\sigma(z)$.

We can use the Fefferman-Graham expansion to compute the boundary energy-momentum tensor
\begin{equation}  
  \label{eq:EMtensor_CSS}
  \begin{split}
    T_{ab} &= \frac{k}{2\pi}\left(g^{(2)}_{ab} - g^{kl}_{(0)} g^{(2)}_{kl} g^{(0)}_{ab}\right)\\
    &= \frac{1}{2\pi}
    \begin{pmatrix}L(z)+\Delta P'(z)^2 & -\Delta P'(z)  \\[2mm] -\Delta P'(z) & \Delta\end{pmatrix}.
  \end{split}
\end{equation}
At this point it should be stated that this energy-momentum tensor is not the canonical energy-momentum tensor for a warped CFT.
For a warped CFT defined on a manifold with warped geometry, the energy-momentum tensor is not symmetric, since symmetry of the energy-momentum tensor is a result of Lorentz invariance.
However, a warped CFT dual to AdS$_3$ with CSS boundary conditions is not defined on a manifold with warped geometry.
Rather, the manifold is not invariant under warped transformations, but for that price we gain the symmetry of the energy-momentum tensor.

The conserved charges corresponding to the asymptotic Killing vectors are
\begin{equation}
  \label{eq:CSS_charges}
  \begin{split}
    Q_{\xi(f)} &= \frac{1}{2\pi}\int_{\pdd \Sigma} d\phi\, n^a T_{ab}\xi^b = \frac{1}{4\pi^2}\int_0^{2\pi} d\phi f(z)L(z) ,\\
    Q_{\eta(f)} &= \frac{1}{2\pi}\int_{\pdd \Sigma} d\phi\, n^a T_{ab}\eta^b = \frac{\Delta}{4\pi^2} \int_0^{2\pi} d\phi f(z) (P'(z)-1),
  \end{split}
\end{equation}
where $\pdd \Sigma$ is at $r\to\infty$, $t = \frac{z+\zb}{2}$ constant, $\phi = \frac{z-\zb}{2} \in (0,2\pi)$, and $n= \pdd_t = \pdd_z+\pdd_\zb$. 

We can now also compute the charge algebra, using the Dirac brackets of Einstein gravity,
\begin{equation}
  \{Q_{\zeta_1(f)},Q_{\zeta_2(g)}\}=\delta_{\zeta_1(f)} Q_{\zeta_2(g) }.
\end{equation}
So we have
\begin{equation}
  \begin{split}
    &\{Q_{\xi(f)},Q_{\xi(g)}\} = \delta_{\xi(f)} Q_{\xi(g)} = \frac{1}{4\pi^2}\int_0^{2\pi} d\phi\, g(z)\delta_{\xi(f)}L(z)\\
    &= \frac{1}{4\pi^2} \int_0^{2\pi} g(z)\left(f(z)L'(z)+2f'(z)L(z)-\frac{k}{2}f'''(z)\right).
  \end{split}
\end{equation}
Expanding the functions in modes,
\begin{equation}
  \begin{split}
    f(z) &=\sum_n f_n e^{inz},\quad g(z) = \sum_m g_m e^{imz},\\
    L(z) &= \sum_p L_p e^{-ipz}.
  \end{split}
\end{equation}
Replacing Dirac brackets with commutators, we obtain the Virasoro algebra
\begin{equation}
  \label{eq:Vir}
  [L_m, L_n]= (m-n)L_{m+n}-\frac{k}{2} n^3\delta_{m,-n}. 
\end{equation}
Note that equating $\frac{k}{2}=\frac{c}{12}$ gives the familiar $c=6k=\frac{3l}{2G_N}$. 
Similarly, we obtain a U(1) Kac-Moody algebra from the commutator of the charges $Q_{\eta}$,
\begin{equation}
  \label{eq:U1km}
  [P_m, P_n] = m \Delta \delta_{m,-n}.
\end{equation}
Note that the Virasoro and Kac-Moody algebra is factorized in this basis. 
This is presented in this form in \cite{Apolo:2018eky}, which also gives the relation between this and the algebra presented in \cite{Compere:2013bya}.

\section{$\ttb$ deformed CSS boundary conditions}
\label{sec:defcss}

To compute the $\ttb$ deformed bulk metric corresponding to the $\ttb$ deformed boundary WCFT, we first compute the deformed boundary metric using \eqref{eq:defrmfg}
\begin{equation}
  \begin{split}
    \gamma_{ij}(\lambda)dz^i dz^j &= -\Big(d\zb+(\lambda L(z) - P'(z))dz \Big)\times\\    
    &\quad\Big(dz + \lambda\Delta(d\zb - P'(z)dz)\Big).
  \end{split}
\end{equation}
This metric is flat, so we express it in explicitly flat coordinates with indices $a,b$, 
\begin{equation}
  \gamma_{ab}(\lambda)du^a du^b = -du\,dv. 
\end{equation}
Equating the two, we can calculate the state dependent coordinate transformation for a $\ttb$ deformed WCFT, analogous to the ones introduced in \cite{Dubovsky:2012wk, Dubovsky:2017cnj}:
\begin{equation}
  \label{eq:sdct}
  \begin{split}
    du &= dz + \lambda \Delta \left(d\zb -P'(z)dz\right),\\         
    dv &= d\zb + dz\left(\lambda L(z) -P'(z)\right),\\
    dz &= \frac{du - \lambda \Delta dv}{1-\lambda^2\Delta L(z)},\\
    d\zb &= \frac{(P'(z)-\lambda L(z))du+(\lambda \Delta P'(z)-1)dv}{1-\lambda^2\Delta L(z)}.
  \end{split}
\end{equation}
Furthermore, we can use the flow equations to compute the full bulk metric dual to the $\ttb$ deformed WCFT
\begin{equation}
  \label{eq:bulklambda}
  \begin{split}
    \frac{ds^2}{l^2} &= \frac{ dr^2}{r^2}+\frac{\left(du \left( \lambda \Delta L+k r^2\right)-\Delta  dv \left(\lambda k r^2+1\right)\right)}{k^2 r^2 \left(\lambda ^2 \Delta L-1\right)^2}\times\\
                     &\left(du \left(\lambda  k r^2+1\right) L-dv \left(\lambda \Delta L+k r^2\right)\right),
  \end{split}
\end{equation}
where $L\equiv L(u,v)=L(z)$. 
Note that on doing so, we lose the $P(z)$ degree of freedom since this is equivalent to imposing Dirichlet boundary conditions at the constant radial surface $r_c = \sqrt{-\frac{1}{k\lambda}}$. 
If we are to impose Dirichlet-Neumann boundary conditions on this surface, we can recover the U(1) degree of freedom.
To do so, we have to perform the transformation
\begin{equation}
  \label{eq:putP}
  u\to u - \lambda \Delta P(u,v),\quad v \to v - P(u,v).
\end{equation}
This is the analogue of the warped symmetry transformation but now in the state dependent coordinates. 
We will explore both types of boundary conditions, starting with the simpler case of only imposing Dirichlet boundary conditions.

\subsection{Asymptotic Killing Vectors I: Dirichlet boundary conditions}

We will first compute the $\ttb$ deformed asymptotic symmetries which preserve the deformed boundary conditions, which is equivalent to imposing Dirichlet boundary conditions at the radial cutoff surface.

Preserving radial gauge \eqref{eq:radialgauge}, we see that the asymptotic Killing vector in the deformed spacetime has the form
\begin{equation}
  \label{eq:radialgaugexilambda}
  \begin{split}
    \xi^r(\lambda) &= r f(u,v),\\
    \xi^u(\lambda) &= V^u(u,v) - \frac{k}{k^2 r^4 - \Delta L}\Big(\Delta(2 \lambda k r^2 + \lambda^2\Delta L+1)\pdd_u f \\
                   & + (\lambda \Delta L(2 + k \lambda r^2) + k r^2)\pdd_v f\Big),\\
    \xi^v(\lambda) &= V^v(u,v) - \frac{k}{k^2 r^4 - \Delta L}\Big(L(2 \lambda k r^2 + \lambda^2\Delta L+1)\pdd_v f \\
                   & + (\lambda \Delta L(2 + k \lambda r^2) + k r^2)\pdd_u f\Big).\\
  \end{split}
\end{equation}

It will be convenient to define
\begin{equation}
  \label{eq:fnshift}
  \begin{split}
    W^u(u,v) &= V^u(u,v) + k\lambda \pdd_\zb f(u,v),\\
    W^v(u,v) &= V^v(u,v) +k\lambda\big( \pdd_z f(u,v) + P'(u,v)\pdd_\zb f(u,v)\big),
  \end{split}
\end{equation}
where, using \eqref{eq:sdct}, the derivatives in $z,\zb$ are
\begin{equation}
  \label{eq:pddz}
  \pdd_\zb = \lambda \Delta \pdd_u + \pdd_v, \quad \pdd_z = \pdd_u  + \lambda L(u,v) \pdd_v - P'(u,v)\pdd_\zb.
\end{equation}

In terms of $W^a$, the mixed boundary condition, or equivalently the Dirichlet boundary condition at $r=r_c$
\begin{equation}
  \mc L_{\xi(\lambda)} g_{\mu\nu} (\lambda) |_{r=r_c}=0,
\end{equation}
constrains the functions in $\xi(\lambda)$ to obey
\begin{equation}
  \label{eq:diric1}
  \begin{split}
    f(u,v) &= -\frac12 \left(\frac{1-\lambda^2 \Delta L}{1+\lambda^2 \Delta L}\right)(\pdd_u W^u +\pdd_v W^v),\\
    W^u &= -\left(\frac{\lambda \Delta}{1+\lambda^2 \Delta L}\right)(\pdd_u W^u +\pdd_v W^v),\\
    W^v &= -\left(\frac{\lambda L}{1+\lambda^2 \Delta L}\right)(\pdd_u W^u +\pdd_v W^v). 
  \end{split}
\end{equation}
It turns out that this is not enough to solve for $\delta L$.
In the undeformed case \eqref{eq:CSSxiL}, the functions in the asymptotic Killing vector are all holomorphic functions, so we apply the holomorphicity property in the deformed case as well:
\begin{equation}
  \label{eq:nozb}
  \pdd_\zb W^a(u,v) = 0,\quad \pdd_\zb L(u,v) = 0. 
\end{equation}
Combining the previous two equations, we get the conditions
\begin{equation}
  \label{eq:diric2}
  \begin{split}
    f(u,v) &= -\frac12 (1-\lambda^2\Delta L(u,v))\pdd_u W^u(u,v),\\
     \pdd_v L(u,v) &= -\lambda \Delta \pdd_u L(u,v),\\
    \pdd_v W^u(u,v) &= -\lambda \Delta \pdd_u W^u(u,v),\\
    \pdd_a W^v(u,v) &= -\lambda L(u,v) \pdd_a W^u(u,v). 
  \end{split}
\end{equation}
We can use these equations to eliminate $v$ derivatives of all the functions, and all derivatives of $W^v$. 

Now we have enough information to be able to solve for $\delta L$.
To do so, we solve 
\begin{equation}
  \mc L_{\xi(\lambda)} g_{\mu\nu} (\lambda) = \pdd_{L(u,v)} g_{\mu\nu}\, \delta_\xi L(u,v). 
\end{equation}
There are three equations but with the relations in \eqref{eq:diric2}, all three equations become identical, and the $r$ dependence drops out.  
Solving for $\delta L$, we get  
\begin{equation}
  \label{eq:deltaLnoP}
  \begin{split}    
    &\delta_\xi L(u,v) =  (W^u-\lambda \Delta W^v) L'\\
    & +\frac12\big(4L+\lambda^2 k \Delta(1-\lambda^2 \Delta L)L''\big)(1-\lambda^2 \Delta L){W^u}'\\
    & + \lambda^2 k\Delta  (1-\lambda^2 \Delta L)^2 L' {W^u}'' - \frac k2 (1-\lambda^2 \Delta L)^3 {W^u}''',
  \end{split}
\end{equation}
where $' = \pdd_u$.
When $\lambda\to 0$, we recover \eqref{eq:dCSSxi}.
Note that this depends on two arbitrary functions $W^u$ and $W^v$.

\subsubsection{Deformed Charge Algebra}

To compute the symmetry algebra of the $\ttb$ deformed holographic WCFT, we must compute the conserved charge algebra of the dual spacetime. 
We first compute the deformed boundary energy-momentum tensor using the flow equations, which also coincides with the AdS$_3$ Brown-York energy-momentum tensor evaluated on the constant radial surface $r=r_c$, 
\begin{equation}
  \label{eq:defstress}
  T_{ij}^{(\lambda)} = -\frac{l}{2\pi}
  \begin{pmatrix}
    \frac{L}{1-\lambda^2 \Delta L} &  \frac{1+\lambda k + \lambda^2 \Delta L}{\lambda^2 k (1-\lambda^2 \Delta L)}\\[2mm]
    \frac{1+\lambda k + \lambda^2 \Delta L}{\lambda^2 k (1-\lambda^2 \Delta L)}&  \frac{\Delta}{1-\lambda^2 \Delta L}\\
  \end{pmatrix}.
\end{equation}

Conserved charges are defined with respect to a constant time coordinate $t$, which is defined in terms of $u,v$  by
\begin{equation}
  u = t + \phi,\quad v = t - \phi. 
\end{equation}
Since $L$ is holomorphic in $z$, we can express the $t$ derivative in terms of the $\phi$ derivative
\begin{equation}
  \pdd_t L = \frac{1+\lambda \Delta}{1-\lambda \Delta}\pdd_\phi L.
\end{equation}
So we can express the $u$ derivatives of holomorphic functions only in $\phi$ derivatives as well
\begin{equation}
  \label{eq:pdutophi}
  \pdd_u  = \frac12 (\pdd_t + \pdd_\phi) = \frac{1}{1-\lambda \Delta} \pdd_\phi. 
\end{equation}
For constant $t$, we can now eliminate $W^v$ in \eqref{eq:deltaLnoP}, using equations \eqref{eq:diric2} and \eqref{eq:pdutophi},
\begin{equation}
  \pdd_u W^v = -\lambda L(\phi) \pdd_u W^u \implies \pdd_\phi W^v = -\lambda L(\phi) \pdd_\phi W^u. 
\end{equation}
Integrating over $\phi$, we have
\begin{equation}
  \label{eq:Wvrep}
  W^v(\phi) = \int^{\phi} d\phi'\, L(\phi') \pdd_{\phi'}W^u(\phi').
\end{equation}
Now we can label the variation of the conserved charges with only one arbitrary function $W^u$. 
Using \eqref{eq:CSS_charges} but with the deformed energy-momentum tensor, the conserved charge is
\begin{equation}
  Q_{f} = \frac{l}{4\pi^2}\int_0^{2\pi} d\phi f(\phi) \frac{\Delta - L(\phi)}{1-\lambda^2\Delta L(\phi)}.
\end{equation}
We can now compute the charge algebra:
\begin{equation}
  \begin{split}
    &\{Q_{W},Q_f\} = \delta_{W}Q_f\\
    &= \frac{l}{4\pi^2}\int_0^{2\pi}d\phi\,f \left(\frac{-\delta_W L}{1-\lambda^2 \Delta L}+\frac{\Delta-L}{(1-\lambda^2 \Delta L)^2}(\lambda^2\Delta \delta_W L)\right).
    \end{split}
\end{equation}
Using \eqref{eq:deltaLnoP} and \eqref{eq:Wvrep},  substituting $f(\phi) = e^{i m \phi},\, W(\phi) = e^{i n \phi}$, and removing $\phi$ derivatives from $L$ using integration by parts, we have
\begin{widetext}
\begin{equation}
  \label{eq:lambdaVir}
  \begin{split}
    &\{Q_{W},Q_f\}=    \delta_{W}Q_f = \frac{l(1+\lambda\Delta)}{8\pi^2(1-\lambda \Delta)^2}\int_0^{2\pi}d\phi\,\frac{1}{1-\lambda^2 \Delta L(\phi)}\Bigg[2 i n^3 \,k (1-\lambda\Delta)^3 e^{i(m+n)\phi}(1-\lambda^2\Delta L(\phi))\\
    &\qquad + e^{i m \phi}L(\phi)\Bigg(2 m n\, \lambda \Delta(1 - \lambda \Delta)^2\int^\phi e^{in\phi'}L(\phi')d\phi' -i e^{in\phi}\big(n\,\lambda^2k\Delta(1-\lambda^2\Delta L(\phi))\\
    &\times \big(n^2\,\lambda\Delta(3-\lambda \Delta(3-\lambda\Delta))-m^2\big)\big) -2(1-\lambda\Delta)^2\big((m-n)-2n\,\lambda \Delta L(\phi)\big)\Bigg)\Bigg].
  \end{split}
\end{equation}
\end{widetext}
Since $L$ is not independent of $t$, only the zero modes are conserved in time.
In this choice of basis of functions and Fourier modes, the central charge term is state dependent.
This is similar to what was found in \cite{Guica:2019nzm} for a $\ttb$ deformed CFT. 
It is straightforward to verify that on taking the $\lambda\to 0$ limit and expressing $L$ in Fourier modes, one recovers the Virasoro algebra.

\subsection{Asymptotic Killing Vectors II: Dirichlet-Neumann boundary conditions}
If we want to impose the same boundary conditions at the radial cutoff in the $\ttb$ deformed metric as the undeformed Dirichlet-Neumann CSS boundary conditions at infinity of the undeformed metric, we have to find a global Killing vector which corresponds to translations on the boundary.
It is easy to verify that $\lambda\Delta\pdd_u +\pdd_v$ is such a global Killing vector of \eqref{eq:bulklambda}.
To generate transformations in the boundary metric, we then promote this global Killing vector to an ``asymptotic Killing vector'' analagous to \eqref{eq:CSSxiP},
\begin{equation}
  \label{eq:etaP}
  \eta(\lambda;\sigma) = -\sigma(u,v)(\lambda\Delta\pdd_u +\pdd_v).
\end{equation}

To introduce the $P(z)$ degree of freedom back into the metric \eqref{eq:bulklambda}, one can make the coordinate transformation \eqref{eq:putP}:
\begin{equation}
  \label{eq:warpeduv}
  u\to u - \lambda \Delta h(u,v),\quad v \to v - h(u,v),
\end{equation}
which is generated by the asymptotic Killing vector \eqref{eq:etaP} as the analogue to the warped transformation $\zb \to \zb - h(z)$.
The state dependent coordinate transformation is now
\begin{equation}
  \begin{split}
    &du - \lambda \Delta \dd(h(u,v))= dz + \lambda \Delta \dd(\zb - P(z)),\\
    &dv - \dd(h(u,v))= d\zb +(\lambda L(z)-P'(z)) dz,\\
    &dz = \frac{du - \lambda \Delta dv}{1-\lambda^2 \Delta L},\\
    &d\zb = \frac{dv - \lambda L du + (du + \lambda \Delta dv)P'(z) }{1-\lambda^2 \Delta L}-(\dd(h(u,v)),
  \end{split}
\end{equation}
where $\dd$ is the exterior derivative.
Note that since both $h$ and $P$ are arbitrary functions of $(u,v)$, we can choose the gauge where $h=P$.
The coordinate transformation now becomes much simpler, 
\begin{equation}
  \begin{gathered}
    du = dz + \lambda \Delta \zb,\quad dv = d\zb + \lambda L(z) dz,\\
    dz = \frac{du - \lambda \Delta dv}{1-\lambda^2 \Delta L}, \quad d\zb = \frac{dv - \lambda L du }{1-\lambda^2 \Delta L}.
  \end{gathered}
\end{equation}
The metric now reads
\begin{widetext}
  \begin{equation}
    \begin{split}
      ds^2 &= l^2 \frac{dr^2}{r^2} + \frac{l^2}{k^2 r^2 (1-\lambda^2 \Delta L)^2}\Big(\big(kr^2(\lambda^2\Delta L-1)\pdd_uh - (1+\lambda kr^2)L\big)du +\big(k r^2 \pdd_vh \left(\lambda^2\Delta L -1\right)+\lambda \Delta L +k r^2\big)dv\Big)\times\\
           &\Big(\big(\Delta  \pdd_uh \left( \lambda^2\Delta  L-1\right)- \lambda\Delta  L-k r^2\big)du+\Delta\big(\pdd_vh \left( \lambda^2 \Delta L -1\right)+\lambda k r^2+1\big)dv\Big).
    \end{split}
  \end{equation}
\end{widetext}

This metric still has the asymptotic Killing vector \eqref{eq:etaP}, and when $\sigma=1$ it is a global Killing vector.
Computing the flow in phase space generated by \eqref{eq:etaP}, we have
\begin{equation}
  \mc L_{\eta(\lambda,\sigma)} g_{\mu\nu}(\lambda; L(u,v), \pdd_u h(u,v)) = \pdd_L g_{\mu\nu} \delta L + \pdd_{h} g_{\mu\nu} \delta h, 
\end{equation}
which on solving, we see that we recover the undeformed U(1) symmetry, 
\begin{equation}
  \delta L = 0,\quad \delta h = \sigma. 
\end{equation}

Now we will see if on performing the warped transformation \eqref{eq:warpeduv} we lose the deformed Virasoro symmetry \eqref{eq:deltaLnoP}.
The vector field which preserves radial gauge is
\begin{widetext}
  \begin{equation}
    \begin{split}
      \xi^r &= r f(u,v),\\
      \xi^u &= V^u(u,v) + \frac{k}{(\Delta L-k^2r^4)(1-\lambda\Delta\pdd_uh-\pdd_vh)^2}
              \Big[\pdd_vf \Big(\Delta  \lambda  L \left(\Delta  \lambda  \pdd_uh \left(-2 \pdd_vh +k \lambda  r^2+1\right)\right.\\
            &\quad\left.+\left(1- \lambda k r^2\right) \pdd_vh - \lambda k r^2-2\right)+\Delta  \pdd_u h \left(\pdd_v h- \lambda k r^2-1\right)+k r^2 \left(\pdd_vh-1\right)+\lambda^3 \Delta ^2  L^2\pdd_vh \left(\lambda\Delta  \pdd_uh-1\right)\Big)\\
            &\quad-\Delta  \pdd_u f\left(2 \left(\lambda k r^2+1\right) \left(\lambda^2\Delta L-1\right)\pdd_vh +(\pdd_vh)^2 \left(\lambda^2\Delta L-1\right)^2+  \lambda^2\Delta L+2  \lambda k r^2+1\right)\Big],\\
      \xi^v &= V^v(u,v) + \frac{k}{(\Delta L-k^2r^4)(1-\lambda\Delta\pdd_uh-\pdd_vh)^2}\Big[ \pdd_uf \Big(-\pdd_vh \left(\Delta  \lambda^2 L-1\right) \left(\Delta  \lambda  L+k r^2\right)\\
            &\quad+\Delta  \pdd_uh \left(\Delta  \lambda^2 L-1\right) \left(\pdd_vh \left(\Delta  \lambda ^2 L-1\right) +k \lambda  r^2+1\right)-\lambda \Delta L \left(\lambda k r^2+2\right) -k r^2\Big)\\ 
            &\quad+\pdd_vf \Big(2 \pdd_uh \left(\lambda ^2 \Delta L-1\right) \left(\Delta  \lambda  L+k r^2\right)-\Delta  (\pdd_uh)^2 \left(\Delta  \lambda^2 L-1\right)^2-L \left(\Delta  \lambda ^2 L+2 k \lambda  r^2+1\right)\Big)\Big].
    \end{split}
  \end{equation}
\end{widetext}
To have solutions which preserve the mixed boundary conditions, we are required to impose holomorphicity of $h$ in $z,\zb$ coordinates. 
\begin{equation}
  \pdd_\zb h = 0 \implies \pdd_v h= -\lambda \Delta \pdd_u h. 
\end{equation}
To simplify the equations, one can introduce the following definitions
\begin{equation}
  \begin{split}
    W^u &= V^u - \lambda^3 k\Delta^2 (1-\lambda^2\Delta L)\pdd_uf \pdd_u h,\\ 
    W^v &= V^v + \lambda k (1-\lambda^2\Delta L)\pdd_uf (1-\lambda\Delta \pdd_u h),\\
    X &= W^u - \lambda \Delta W^v. 
  \end{split}
\end{equation}
To compute the variation of the functions $L$ and $h$, it will be necessary to impose holomorphicity in $z$ for the functions $W^a, X, f$.
We can now compute the variation of the metric which preserves the mixed boundary conditions, or equivalently impose Dirichlet boundary conditions at the constant radial surface
\begin{equation}
  \mc L_{\xi(\lambda,\sigma)} g_{\mu\nu}(\lambda; L(u,v), \pdd_u h(u,v)) |_{r=r_c} = 0. 
\end{equation}
This is a set of three equations, however only two are linearly independent.
The conditions we get from solving the above equations are
\begin{equation}
  \begin{split}
    f(u,v) &= -\frac12 \left(\frac{1-\lambda^2\Delta L}{1+\lambda^2 \Delta L}\right) X'(u,v),\\
    X(u,v) &= \frac{W^v(u,v)}{h''(u,v)}\\
           &\; - \frac{h'(u,v) - \lambda L (1-\lambda \Delta h'(u,v))X'(u,v)}{h''(u,v)(1+\lambda^2\Delta L)},
  \end{split}    
\end{equation}
where $' \equiv \pdd_u$. 

We are now in a position to compute the flow of the metric in phase space generated by this vector field subject to the above constraints,
\begin{equation}
  \mc L_{\xi(\lambda,\sigma)} g_{\mu\nu}(\lambda; L(u,v), \pdd_u h(u,v)) = \pdd_L g_{\mu\nu} \delta L_\xi + \pdd_{h} g_{\mu\nu} \delta h_\xi. 
\end{equation}
As before, this set of three equations, subject to the constraints reduce to two equations, and remove any dependence on the radial coordinate $r$.
The variation of the functions $h$ and $L$ are 
\begin{widetext}
  \begin{equation}
    \begin{split}
      \delta_\xi h &= 0,\\
      \delta_\xi L &= \frac{1}{2 \Theta^2 h''} \Bigg(\Big(2\lambda k L' L_m^2 L_p (2\lambda \Delta h' -1)h'' + 3 k L_m^3 L_p^2(h'')^2\Big)W'' - k\Theta L_m^3 L_p h'' W''' \\
                   &\qquad + X' \Big(-2 \lambda^2 k (L')^2 L_m^2 h''\big(1-3\lambda\Delta h' + 2\lambda^2\Delta^2 (h')^2\big) - 6 k L_m^3 L_p^2 (h'')^3 +h''\big(-\Theta L_m (\lambda k L''(1-2\lambda \Delta h') \\
                   &\qquad + 4 \lambda L^2 (1-\lambda\Delta h') + L(-\lambda^3 k \Delta L'' -2(2-\lambda^4 k \Delta ^2L'')h')) +6k \Theta L_m^3 L_p h'''\big)-L'\big(2 \Theta^3 \\
                   &\qquad  + \lambda k L_m^2(-7+ 8\lambda \Delta h' - \lambda^2 \Delta L (1-8\lambda \Delta h'))(h'')^2+2 \lambda k \Theta L_m^2(1-2\lambda\Delta h')h'''\big)- k \Theta^2 L_m^3 h''''\Big) \\
                   &\qquad + W'\Big( L'\big(2 \Theta^2 L_p + 2 \lambda k L_m^2 L_p(1-2\lambda \Delta h')h'''  + k \Theta L_m^3 L_p h''''\big)-3k L_m^3 L_p^2 h'' h'''\Big)\Bigg),
    \end{split}
  \end{equation}

\end{widetext}
where 
\begin{equation}
  \begin{gathered}
    \Theta = \pdd_u h - \lambda L(1 - \lambda \Delta \pdd_u h),\quad  W = W^v,\\
    L_m = 1-\lambda^2 \Delta L,\quad L_p = 1+\lambda^2 \Delta L.
  \end{gathered}      
\end{equation}
We see that we still preserve a deformed Virasoro generator, and do not generate a transformation in the U(1) generator.
However, the algebra produced the modes of the charges will not be closed as the variation of $L$ depends on the U(1) generator $h$.
We will not compute a charge algebra for this since it is not illuminating, but in principle can be computed using the same procedure outlined in the previous section. 

\section{Discussion}
\label{sec:disc}
In this paper we computed the $\ttb$ deformed generators of a Warped CFT, using holographic techniques developed in \cite{Guica:2019nzm, Kraus:2021cwf}.
Previously, holographic $\ttb$ techniques have been used to compute $\ttb$ deformations of holographic CFTs.
Since the  $\ttb$ deformation is a double trace deformation, the boundary conditions of the holographic bulk dual are modified.
For the $\ttb$ deformation, this can be interpreted as imposing Dirichlet boundary conditions at a finite radial surface for the bulk metric.
However, when considering a holographic WCFT dual to AdS$_3$, one has to employ the CSS boundary conditions \cite{Compere:2013bya}, which are Dirichlet-Neumann boundary conditions for the bulk metric.

We therefore computed the $\ttb$ deformed CSS boundary conditions, by imposing either only Dirchlet boundary conditions at the radial surface or Dirichlet-Neumann boundary conditions at the same surface.
Using this, we computed the $\ttb$ deformed asymptotic symmetry algebra for both cases, and found that for a $\ttb$ deformed holographic WCFT, the U(1) Kac-Moody generators are not affected, but the Virasoro generators are deformed in a non-linear way.
In fact, when considering the Dirichlet-Neumann boundary conditions at the finite radial surface, we see that the deformed Virasoro generator will no longer create a closed algebra.
This suggests that the symmetry algebra of the $\ttb$ deformed WCFT still contains the U(1) Kac-Moody algebra, which follows from the fact that the $\ttb$ deformation preserves translation invariance. 

This result strengthens and extends the proposals of \cite{McGough:2016lol,Guica:2019nzm,Kraus:2021cwf} to the case of an example of bottom-up holography where the boundary theory is not a conformal field theory, but instead a non-relativistic theory. 
It will be interesting to explore how the holographic $\ttb$ dictionary extends to other examples of holography, and in particular, non-AdS holography. 

There are many directions one can take from here.
Another starting point for a bulk dual to a non-relativistic QFT would be the a $J\ovl T$ deformation of a CFT dual to AdS$_3$ with a U(1) Chern-Simons matter field, to generate CSS-like boundary conditions \cite{Bzowski:2018pcy}.
Since Warped CFTs can also be formulated as dual to modified gravity theories with a Warped AdS bulk, it would also be interesting to use the Chern-Simons formalism of holographic $\ttb$ \cite{Llabres:2019jtx} to compute the $\ttb$ deformations of WCFT dual to Warped AdS$_3$ as a solution to Lower Spin Gravity \cite{Hofman:2014loa}, or as a solution to Massive Gravity \cite{DESER1982372,Bergshoeff:2009hq}.
Stepping away from holography, it would be interesting to compute the $\ttb$ deformed WCFT partition function, and explore the deformations of other non-relativistic QFTs such as the Quantum Lifshitz Model in 2+1d, which will require understanding $\ttb$ deformations in higher dimensions. 

\begin{acknowledgments}
  The author would like to thank Valentina Giangreco M. Puletti, Monica Guica, and L\'arus Thorlacius for their insights and helpful discussions.
  This work was supported by the Icelandic Research Fund under the grant 228952-052 and by a University of Iceland Doctoral Grant.
\end{acknowledgments}

\bibliography{warpedbib}

\end{document}